# Block Copolymer Double Diamond Twin

Xueyan Feng[1,2,*], Michael S. Dimitriyev[3] and Edwin L. Thomas[1,*]

1. Department of Materials Science and Engineering, Texas A&M University, College Station, TX, USA
2. State Key Laboratory of Molecular Engineering of Polymers, Department of Macromolecular Science, Fudan University, Shanghai, China
3. Department of Polymer Science and Engineering, University of Massachusetts, Amherst, MA, USA

**E-mail:** fengxueyan@fudan.edu.cn, elt@tamu.edu

**Abstract**

A twin boundary (TB) is a common low energy planar defect in crystals including those with the atomic diamond structure (C, Si, Ge, etc.). We study twins in a self-assembled soft matter block copolymer (BCP) supramolecular crystal having the *double* diamond (DD) structure, consisting of 2 translationally shifted, interpenetrating diamond networks of the minority polydimethyl siloxane block embedded in a polystyrene block matrix. The coherent, low energy, mirror-symmetric double tubular network twin has one minority block network with its nodes offset from the (222) TB plane while nodes of the second network lie in the plane of the boundary. The offset network, although at a scale about a factor of $10^3$ larger, has precisely the same geometry and symmetry as a (111) twin in atomic single diamond where the tetrahedral units spanning the TB retain nearly the same strut (bond) lengths and strut (bond) angles as in the normal unit cell. In the DD, the second minority block network undergoes a dramatic restructuring – the tetrahedral nodes transform into 2 new types of mirror-symmetric nodes (pentahedral and trihedral) which alternate and link to form a hexagonal mesh in the plane of the TB. The collective re-organization of the supramolecular packing highlights the hierarchical structure of ordered BCP phases and emphasizes the remarkable malleability of soft matter.

**Introduction**

The diamond crystal is one of the most appealing structures in material science. Indeed, the publication of the diamond crystal structure by the Braggs in 1913 can be considered as the start of X-ray crystallography.[1] The structure is composed of a basic unit of 4 tetrahedrally coordinated atoms having sp³ orbital bonding in the $Fd\bar{3}m$ space group. These bonds form a *6⁶-a* network[2] where the smallest closed loop has 6 nodes with each node having four equivalent nearest neighbors (f = 4). The Carbon group elements (i.e. carbon, silicon, germanium) can all display this structure. Single diamond (SD) structures are also observed in systems with much larger building blocks, such as patchy colloidal particle systems[3] and the DNA origami cage/gold nanoparticle system[4]. Many alloys (e.g. beta silicon-carbon, gallium-arsenide, indium-antimonide, etc.) also adopt a similar structure but now two different types of atoms occupy two different Wyckoff sites within the SD network (the so called zincblende structure with space group $F\bar{4}3m$)[5]. The double diamond (DD) structure consists of two independent but interpenetrating diamond networks as occurs in cuprite ($Cu_2O$ with space group of $Pn\bar{3}m$)[6]. Interpenetrating diamond networks are also found in binary alloys. These are known as the NaTi or B32 structure and still retain the diamond $Fd\bar{3}m$ space group since each diamond network is composed of a single type of atom.[7]

For the double diamond structure with both networks made of the same material, there are two equivalent node positions in the primitive cell located at Wyckoff site 2a with tetrahedral point group $\bar{4}3m$ site symmetry (we use the Hermann-Mauguin notation for point groups). Each node has four connections (f = 4) to nearest neighbor nodes. Continuous tubular SD and DD network structures can also form at the mesoscale in soft matter systems, including microemulsions[8], liquid crystals[9], block copolymers (BCPs)[10], and even insect scales[11]. A tubular DD network was first identified in 1983 in the glycerol monooleate/water system[8], while in BCP systems, both $Pn\bar{3}m$ (AB linear diblock)[10] and $Fd\bar{3}m$ (ABC linear triblock)[12] variants have been observed.

Due to their 3D connectivity and symmetries, materials possessing either SD or DD continuous tubular network structures (with a continuous matrix separating the networks, thus a bi/triply continuous structure) can exhibit interesting behavior such as ion or electron or phonon transport, mechanical properties, and photonic crystal band gaps[13,14]. Unlike atomic crystals, where the centers of the various types of atoms are precisely located at specific Wyckoff positions, the 3D crystal periodicity of soft matter, such as the supramolecular assembled BCP diamond tubular networks, arises from the shape and translational periodicity of (sets of) triply periodic soft interfaces. These interfaces, called the intermaterial dividing surfaces (IMDS)[15] are regions where the composition changes from minority block A to majority block B. In the case of block copolymers, the basic unit cell is a collection of thousands of molecules containing millions of atoms, and due to the noncrystalline nature of the glassy and/or rubbery component blocks, the molecules have considerable positional and orientational freedom. There are only relatively mild restrictions with the localization of the A block- B block junction at the IMDS and the requirement for both blocks to uniformly fill space. Here, a tetrapodal minority network subdomain, surrounded by the outer majority matrix subdomain, forms a basic motif in the DD networks, designated as a "mesoatom" (Figure S1)[16]. Due to the relatively weak physical interactions between the molecules, block copolymer "mesoatoms" as well as their associated microdomain IMDS shapes are much more malleable than their atomic counterparts and can adjust their dimensions and shape in response to the presence of defects or other sources of stress.[17,28]

Investigating the types of defects that occur within a material and their influence on the properties of the material is a central theme of materials science.[18] Designing and incorporating desirable defects into a material (e.g. grain boundary engineering) can provide tailored properties (e.g. thermal conductivity, creep resistance).[19-21] While grain boundaries in hard matter have been the subject of investigation for over a century, corresponding grain boundary studies in soft matter, such as block copolymer assemblies, are relatively few.[22-29]

Twins are very common planar defects in crystals. Unlike point defects such as vacancies and impurity atoms, line defects such as dislocations and disclinations, and general grain boundaries, all which break and reduce symmetry, the appearance of a TB *introduces* an additional symmetry element (reflection or rotation or inversion) which did not previously exist at the original position in the crystal structure[30]. TBs have been noted in a vast number of materials ranging from inorganic hard matter[30-33] to organic soft matter[34,35]. The fine scale structure of twins has been studied with high-resolution transmission electron microscopy (HRTEM) and diffraction techniques, allowing for determination of the atomic structure of the TB for many hard crystal materials, including those with diamond symmetry[36-40]. TBs in SD structures occur on (111) planes (based on origin choice 2 with an inversion center at the unit cell origin for space group $Fd\bar{3}m$) and can appear during crystal growth or due to stresses arising in phase transformations or from mechanical loading[41,42]. HRTEM and electron diffraction studies of the (111) diamond twin show that the TB does not alter any atomic bond lengths or bond angles resulting in a low energy defect[36,43]. Moreover, twins can alter the macroscopic properties of a material: diamond crystals with a high density of lamellar twins exhibit superior hardness[44].

While the precise structural details of TBs in atomic diamond structures are well studied and understood, the nature of TBs in mesoscopic scale soft matter diamond structures has only been addressed in the past few years.[45-48] Comparison of image simulations of various twin boundary models with the TEM images suggest a {111} type twin in cubosomes (see *Prior Imaging of Diamond Twins in Soft Matter* section in the SI for further discussion). A recent investigation has confirmed the existence of triply periodic minimal surfaces (TPMS) with a mirror boundary ("minimal twin surfaces"), including the {111} twin in Schwarz's D surface.[49] Thus, while there is a growing literature concerning mesoscale twins occurring in DD soft matter crystals and mathematical justification for the smooth incorporation of a {111} twin feature into a minimal surface, there is, as yet, no direct 3D experimental evidence of the actual structure of a TB in a DD soft matter crystal.

Slice-and-view scanning electron microscopy (SVSEM) is a 3D tomography method that combines focused ion beam (FIB) slicing and scanning electron beam imaging to produce high fidelity, large volume 3D reconstructions, making it possible to unambiguously analyze the complex, extended defect features occurring in soft matter systems. Our group adopted SVSEM for the study of BCP morphology[17,29,50] including detailed characterization of a topological mirror defect (twin boundary) in the double gyroid tubular network structure. Here, using SVSEM on a high molecular weight BCP with the double diamond structure, we identify and characterize the structure of TBs in a polystyrene-*b*-polydimethylsiloxane (PS-PDMS) tubular network at both the supra-unit cell level and sub-unit cell level. Real space and reciprocal (Fourier) space data unequivocally reveal the boundary plane is a (222) mirror plane. Moreover, clear structural details of the 3D geometry, topology, node functionality and new types of mesoatoms of each network, as well as the new symmetries and shapes of the IMDS across the TB, are presented.

**Results and discussion**

A secondary electron image of a Ga$^+$ ion milled surface of a bulk PS-PDMS sample exhibiting alternating periodic patterns is shown in Figure 1a. Adjacent patterns abruptly transform across a set of three sharp, straight, parallel boundaries separating a group of 4 grains. The image alternately shows near 2D {410}, {223} DD patterns within each grain (see simulated 2D patterns in Figure S2). Due to the stronger secondary electron emission from the higher atomic number Si atoms in the PDMS block, the PDMS domains appear bright while the PS domains appear dark. SVSEM tomographic reconstruction on a region containing one of the boundaries is shown in Figure 1b. After binarizing the grey scale data based on the volume fractions of the PS-PDMS material (42% PDMS), the 3D reconstruction reveals two independent PDMS networks (rendered red and blue, each enclosing 21% volume fraction, while the PS matrix is rendered transparent (58% volume fraction). The sharp orange line in Figure 1c indicates part of the boundary plane between adjacent grains. Software-based slicing of the 3D data along the direction normal to the 2D boundary plane yields a series of 2D periodic patterns very similar to slicing along the [111] direction of a level set approximation to the DD structure[51] with 42% minority component (see Supporting Video 1). The boundary plane indexes to the (222) based on unit cell origin choice 2 for $Pn\bar{3}m$ (Figure 1b). The boundary provides mirror symmetry between the grains as evident from the colored network reconstructions embedded within the more transparent larger reconstruction (Figure 1c).

Selected volume diffraction[17,28] was used to determine the crystallographic directions and to quantify the unit cell parameters for regions in grain 1 and grain 2 adjacent to the boundary (Figure S3). To index the boundary plane with the same Miller indices from both grains, we employ left-handed unit cell coordinates for grain 1 and right-handed unit cell coordinates for grain 2. The mirror coherency between grain 1 and grain 2 can also be clearly identified from selected 2D Fourier sections of the 3D volume diffraction patterns. As shown in the $(1\bar{1}0)^*$ section (Figure 1d), where * indicates reciprocal space, the grains have colinear $[111]^*$ and $[\bar{1}\bar{1}\bar{1}]^*$ vectors and the Bragg peaks for grain 1 and those for grain 2 are distributed in a mirror-symmetric way with the mirror line along $[11\bar{2}]^*$.

As an aid for analyzing the geometry, topology and IMDS shape of the experimentally observed (222) DD twin, we construct a simple model (see Figure S4). An initial level set DD tubular network model (42% volume fraction of minority networks) is made and terminated with a (222) surface plane. Note that the structure of each of the two PDMS networks is quite different on the terminating surface. To form the twin model, a second level set structure is produced by mirror reflection across the terminating (222) plane of the first structure. Fusion (without modification or smoothing of the IMDS) of these two volumes across their terminating (222) surface planes creates a 3D model of the (222) TB volume.

Detailed comparison of the experimental 3D reconstructed structure of the BCP sample and the level-set (222) TB model structure are shown in Figure 1c, 2a and Figure S4. Both the experimental and model red network display the same structure with no nodes on the TB and the TB plane merely changing the symmetry between the pair of [111] boundary spanning f = 4 nodes from inversion symmetry into mirror symmetry. Whereas, the experimental blue network reveals major adjustments in the local geometry, topology and shape of the IMDS compared to the model. As evident in Figures 2 and 3, new types of nodes with 3 and 5 struts are formed on the TB, with each type displaying $\bar{6}m2$ point group symmetry (as confirmed through measurements of the IMDS surface normal vector distribution, averaged separately over $f = 3$ and $f = 5$ nodes, see Figure 2(d)). These boundary nodes interconnect to form a 6-node hexagonal mesh structure (Figure 2a). Each 6-node loop of the blue network on the boundary is catenated with a strut from a 6-membered red loop passing through its center. The set of symmetries at the f = 3 and f = 5 boundary nodes consists of the horizontal TB mirror plane, a vertical 3-fold axis along [111], 3 vertical intersecting mirror planes and 3 horizontal intersecting 2-fold axes along <110>. The 2D plane group of the boundary can be identified as p3m1 (Figure 2b). These twin boundary symmetries coincide with the surrounding symmetry elements of the $Pn\bar{3}m$ structure and of course add the additional horizontal mirror symmetry. Note that in the previous 2D TEM and model-based studies of DD twin cubosomes,[47,48] the suggested twin boundary models also had a (222) habit plane with air channels nodes with $\bar{6}m2$ point group symmetry.

A single diamond network is comprised of 6-$4^6$ loops (the m–$f^n$ notation indicates the smallest closed loop consisting of m nodes with each node having f connected neighbors and the exponent n indicating the number of repeats in the loop). We can identify 3 new types of boundary loops in the TB volume (Figure 3a). In the red network, a 6–$4^6$ loop (6 nodes in the loop and each node has 4 connections), spans across the (222) TB. The mirror cuts through two [111] oriented struts of the loop, dividing the loop into two equal, mirror-symmetric parts. Regular diamond 6–$4^6$ loops do not have such mirror symmetry. The second type of boundary loop lies in the blue network within the TB plane, having 6 nodes with alternating f = 5 and f = 3 nodes, 6-$(5,3)^3$. The third type of boundary loop is also in the blue network and spans across the TB in the form of a fused loop pair (U2-6-5,$4^3$,5,3, where U2 means the union of two loops). Here, the TB plane bisects the fused struts common to both loops.

Construction of the skeletal graphs of the DD tubular networks allows for simplification of the structure for comparison with the atomic diamond structure in hard crystals (Figure 3b). The vertices and the edges of the DD skeletal graph correspond to the atomic positions and bond lengths in a diamond crystal. Except for a difference in scale of the order of $10^3$, the vertices and edges of the skeletal graph of the red network in the vicinity of the TB are remarkably similar to the positions of atoms and bonds observed in (111) TBs of atomic diamond[52]. However, the presence of the twin necessitates significant changes in the blue network and hence to its skeletal graph. As previously noted, the blue network reorganizes to form two new types of boundary nodes, struts of different lengths and orientations and two new types of loops to create the mirror-symmetric boundary.

Quantification of the structural features in the red and blue networks as a function of the distance along the perpendicular direction from the TB (i.e. [111] direction) is shown in Figure 4. At the TB, the angle between adjacent struts normal to the TB is 90° (along [111]) and between adjacent struts in the plane of the TB is 120° (along <110>) (see blue scheme in Figure 4b). The inter-strut angle becomes approximately tetrahedral (109.5°) immediately upon leaving the TB (Figure 4c). The volume of an average $f = 5$ mesoatom is $\sim 1.5 \times$ that for the normal $f = 4$ mesoatom while that of the average $f = 3$ mesoatom is $\sim 0.70 \times$ (Figure 4d). Tetrahedral mesoatoms exhibit a small volume increase approaching the TB. The IMDS surface area ($S$) to minority block volume ($V$) ratio ($S/V$) of the three types of mesoatoms shows an opposite trend with the trihedral mesoatoms having the greatest and the pentahedral mesoatoms the least S/V (Figure 4e). The strut lengths within and normal to the TB in the blue network also show interesting variations compared to the average tetrahedral node to tetrahedral node strut distance of $\sim 78$ nm. The pentahedral to trihedral <110> struts in the boundary loop display the shortest length, $\sim 70$ nm (Figure 4f), while the [111] struts that connect the pentahedral nodes in the TB elongate to $\sim 90$ nm to reach their tetrahedral neighbors (Figure 4g). For the red network, the only boundary strut is perpendicular to the TB plane (see red scheme in Figure 4b). As evident in Figure 4h, this strut length is only slightly longer ($\sim 4$-9%) than a normal DD network strut.

The fact that the DD TB supports such network polymorphism and maintains consistent symmetry is remarkable. Recent results concerning triply periodic length-minimizing networks[53,54], skeletal network analogues to TPMSs, show that the gyroid and the diamond are the optimal (per unit cell volume) length-minimizing networks for fixed functionality of $f = 3$ and 4, respectively. For $f = 5$, the optimal network is the bnn (i.e., named after boron nitride nanotube networks)[54], a skeletal relative of one of Schwarz's H surfaces, consisting of planar hexagonal rings of $f = 5$ nodes stacked in parallel layers. The similarity of the TB seen here, as well as the H surface mentioned in Han, et al.[55], with the bnn network is striking, differing in the replacement of every other $f = 5$ node with an $f = 3$. The proximity of the TB network structure to a length-minimizing network structure may point towards one reason for the apparent thermodynamic robustness of the DD TB, despite the required $f = 5$ nodes. The trigonal bipyramidal arrangement of $f = 5$ node strut orientations is consistent with the symmetry requirements of edges in length minimizing networks[56]. Moreover, the variability in strut lengths about the TB suggests an effective force law between nodes that is independent (or weakly dependent) on strut length, in contrast with the observed uniformity of bond lengths in atomic TBs, further supporting the analogy with length-minimizing networks for which all struts are under equal tension regardless of length. Taken together, we conjecture that the skeletal structure of the TB may be a length-minimizing network under the constraints of joining the skeletal networks of the neighboring diamond grains. Regardless, the large variation in strut (bond) length is also associated with the variety in the surface area to volume ratios $S/V$ of nodes in the TB region, as suggested by the spherical scaling relation $S/V \sim L^{-1}$, for length scale L representing the radius of a nodal region, which is bounded by the lengths of the struts emerging from each node. This highlights the way in which broken symmetry in soft crystals can lead to new and diverse organizing environments for the constituent molecules.

The (222) TB studied here is very likely a growth twin. Once formed, the concave re-entrant grooves[57] of the low energy growth faces of the twinned crystal would favor nucleation that stabilizes continuous growth of the twin. In silicon, the re-entrant growth face allows simultaneous bonding of all four sp³ bonds of a Si atom joining the crystal, so that once established, the reentrant growth face promotes rapid addition of atoms. For the mesoscale DD BCP structure, the situation is more complex due to the interpenetrating networks and the need for the formation of the trihedral and pentahedral mesoatom units presumably transformed from the normal tetrahedral units. Future SVSEM imaging involving in-situ examination of a growth front of the DD phase may resolve this question.

**Conclusion**

The 3D structure of a (222) twin in the double diamond phase of a PS-PDMS block copolymer was unambiguously determined using slice-and-view SEM tomography. One network exhibited the same structural symmetry on crossing the TB as a (111) twin in atomic Carbon group single diamond network structures. For the soft matter double diamond twin, a portion of the second network lies on the TB plane and two new types of boundary nodes along with two new types of network loops are formed. Imaging of many structural units in and around the boundary affords statistically sound reconstruction of the many boundary features. The IMDS surrounding the trihedral and pentahedral boundary nodes exhibits $\bar{6}m2$ symmetry and the cross section of the IMDS in the 2D TB plane exhibits plane group p3m1 symmetry. The DD TB structure is an example of network polymorphism with the appearance of $f = 5$ nodes resembling the bnn length-minimizing network but with the replacement of every other $f = 5$ node with an $f = 3$ node likely due to the catenation of the double networks. The influence field of the TB on the DD structure is quite localized as demonstrated by the statistical measurement of key network morphology features such as strut lengths and inter-strut angles using the experimentally determined skeletal graphs in the boundary neighborhood, suggesting a low energetic cost of this type of twin boundary. The malleability of the component mesoatoms constituting the BCP DD tubular networks is key to providing a low energy defect structure. The extensive quantitative 3D structural data available from our study may motivate additional theoretical investigations of malleable self-assembled soft crystal mesostructures concerning their ability to re-structure to accommodate the misfit imposed by a defect[58], the resemblance of features of the TB defect to length-minimizing networks as well as epitaxial relationships in order-order phase transformations[27,59-65].

**Acknowledgements**

The primary support for this work is through a grant to E.L.T. from the National Science Foundation under award DMR 2105296. M.D. was supported by a U.S. Department of Energy (DOE) grant to UMass under award DE-SC0022229. We thank A. Avgeropoulos for providing the block copolymer sample and RM Ho for previous research on network structures with this BCP. V. Subramanian kindly furnished the simulated 2D SEM images in Figure S2.

**Methods**

**Material and sample preparation** The polystyrene-*b*-polydimethylsiloxane (PS-PDMS) diblock copolymer was synthesized by sequential anionic polymerization of styrene and hexamethylcyclotrisiloxane as reported in Ref. 10. The number average molecular weights of the PS block and the PDMS block are 51 kg/mol and 35 kg/mol, respectively. The volume fraction of PDMS block is ~ 42% and dispersity for the whole PS-PDMS polymer is 1.05. The sample studied was slowly cast over 2 weeks from a 5 wt % solution (2ml) in chloroform. The sample was characterized by synchrotron X-ray at 12-ID of National Synchrotron Light Source II in Brookhaven National lab before study by SVSEM. Based on the SAXS pattern, the structure can be nominally associated with a double-diamond morphology, with an average cubic repeat of <*D*> = ~90 nm. Before SVSEM imaging, the sample was attached to a 45º SEM stub with double sided conductive carbon tape; and the outer surface was then coated with a 50 nm layer of platinum.

**Slice-and-view SEM data** The acquisition of SVSEM data follows the procedure reported in Ref. 17. A Thermo Fisher Helios NanoLab 660 SEM-FIB DualBeam system using a gallium ion beam ($Ga^+$) having an energy of 30 keV with a current of 80 pA was used to mill the sample surface. A 1 keV electron beam with beam current of 50 pA was used to image the sample surface with a Through Lens (TLD) secondary electron (SE) detector. Notably, the stronger scattering from the higher atomic number of Si atoms in the PDMS and the resulting additional SE emission is sufficient to provide excellent intrinsic contrast between the PS and PDMS domains without staining. Fiducials were used to register the FIB and SE images during the automatic slice and view process. The pixel resolution for each 2D SEM image collected is 3.58 nm/pixel. The actual FIB slice thicknesses during the image acquisition process were monitored based on FIB images as 2.88 $\pm$ 0.01 nm/slice (details see Figure S5 in SI).

**Data processing and analysis** ImageJ (https://imagej.nih.gov/ij/), was used to binarize the greyscale image stack data to identify the tubular networks formed by the minority domains and separated them from the majority block filled matrix by using a threshold such that the volume fractions of the two binary components matched with the experimentally reported block volume fraction (~42%). Reconstructions and morphological analysis were done with Avizo software from Thermo Fisher and Supporting Software reported in Ref. 17. The supporting software can be found on the online data repository at https://doi.org/10.7275/wv24-3j62.

**3D FFT analysis of grains.** The coherence of the unit cells within a DD grain enables high-resolution analysis of 3D SVSEM data set in reciprocal space, extracting the unit cell vector information. Before transforming the real space volume data into Fourier space, a Hanning window was applied on the raw SEM data volume in order to reduce artifacts in the FFT associated with sample boundary discontinuities. The unit cell vectors of the DD structure can be obtained from the indexed 3D FFT pattern.

**3D reconstruction visualization.** Based on the binarized images, a 3D volume can be reconstructed as colored PDMS networks within a transparent PS matrix using Avizo software from Thermo Fisher. Sub-volumes of regions of interest (ROI) can be further cropped out for local analysis.

**3D Data Treatment for statistical analysis**

IMDS construction: To construct 3D meshed IMDSs from SVSEM data, we followed the procedure outlined in Ref. 17 and adapted the supplemental software. A summary of the process is as follows: (1) Apply a Gaussian blur to the SVSEM density data. (2) Select a trial level set value for the smoothed data as a guess for the location of the IMDS. The initial trial level set value is 0.5. (3) Construct a trial meshed IMDS in Matlab using the isosurface function. (4) Calculate the volume enclosed by the trial surface. If this volume divided by the total volume is within a tolerance of $10^{-3}$ of the PDMS volume fraction (here, 0.42), then the surface is accepted. Otherwise, vary the level set value and repeat steps 2-4.

Skeletonization of density data: The same reference also provides a method for finding the network skeletons. In summary: (1) Using the blurred SVSEM density data, use ImageJ software to skeletonize the individual z-slices of the 3D data. (2) Construct a 3D skeleton by merging nearby points (within a cutoff radius) until only $f = 3, 4$ and 5 skeleton nodes remain. Note that only edges reaching the boundary of the 3D volume are allowed to terminate at nodes of smaller coordination. (3) Using Matlab, vary the positions of the non-boundary nodes until the integral of the density field over all edges is maximized. This is done to make sure that skeleton edges align near the center of the PDMS density field, ensuring that the skeletonization faithfully encodes both the topology and geometry of the double networks.

Skeletal network analysis: The position and orientation of the TB was obtained by fitting a plane to the positions of the $f = 3$ and 5 nodes within the TB. If $\{r_i\}$ are the positions of each node in the TB, then the center of the plane was set as the average over all node positions, $\bar{r} = n^{-1} \sum_{i=1}^{n} r_i$, where $n$ is the number of nodes within the TB. The normal vector $\widehat{N}$, which we use to define an orientation for the [111] direction, was then set along the direction of the eigenvector with minimal eigenvalue of the second moment matrix of node positions, $\sum_{i=1}^{n} r_i \otimes r_i$. Using this fit, we determined the distances of each $f = 4$ node from the TB plane for use in Fig. 4b-d. The [11$\bar{2}$] was then fixed as the average bond direction between alternating nodes in the TB.

Node morphology characterization: We isolated the IMDS of individual nodes by defining a set of cutting planes for each node. These cutting planes were chosen to be centered at the centroid of edges joining the node to its neighbors in the skeletal network and orthogonal to these edges. Mesh faces lying inside the region defined by the cutting planes are included as part of the IMDS belonging to the individual node; mesh faces cut through by a cutting plane were subdivided so that only the portion lying in the cutting plane region counted as part of the nodal IMDS. This was repeated for each node separately, resulting in a disjoint collection of meshed IMDSs. To calculate the total area $S$ for an individual node, we simply summed over the individual areas of each mesh face $S_\alpha$, where $\alpha$ indexes each face in the mesh. The volume of a node was found by first closing off the IMDS with planar polygons glued to the boundaries formed by the mesh cutting operation. Treating these additional polygons as new mesh faces, the volume of the interior of the closed IMDS is $V = \frac{1}{3} \sum_\alpha S_\alpha \widehat{n}_\alpha \cdot (r_\alpha - r^*)$, where $\widehat{n}_\alpha$ is the unit normal to an individual mesh face, $r_\alpha$ is the centroidal vector of the face, and $r^*$ is a point inside the nodal region (we use the position of the corresponding node on the skeletal network).

**Heat maps of the IMDS normal distribution function.** For each nodal IMDS, we calculated a discrete approximation to the normal distribution function $\rho(\widehat{n})$ as detailed in Ref [66]. This discrete approximation is obtained from the meshed nodal IMDSs, from which we can obtain a normal vector $\widehat{n}_\alpha = (\sin \theta_\alpha \cos \varphi_\alpha, \sin \theta_\alpha \sin \varphi_\alpha, \cos \theta_\alpha)$ and area $S_\alpha$ for each face $\alpha$ in the mesh, where $\theta_\alpha$ and $\varphi_\alpha$ are the spherical coordinates representing the orientation of $\hat{n}_\alpha$ relative to the fixed global coordinate system. The normal distribution function is then given by $\rho(\theta, \varphi) = \sum_\alpha [S_\alpha \delta(\theta - \theta_\alpha)\delta(\varphi - \varphi_\alpha)]/(S \sin\theta)$, where $S = \sum_\alpha S_\alpha$ is the total mesh area. To convert this discrete distribution into the continuous distribution of Fig. 2d, we use a truncated spherical harmonic expansion, $\rho(\theta, \varphi) = \sum_{\ell=0}^{\ell_c} \sum_{m=-\ell}^{\ell} \bar{C}_\ell^m Y_\ell^m(\theta, \varphi)$, where $Y_\ell^m(\theta, \varphi)$ are spherical harmonics, $\ell_c = 10$ is the chosen cutoff for the expansion, and $\bar{C}_\ell^m$ are the coefficients of the expansion, obtained as an average of individual nodal values $C_\ell^m = \sum_\alpha Y_\ell^{m*}(\theta_\alpha, \varphi_\alpha) S_\alpha/S$ over each sub-population of nodes (either the $f = 3$ and 5 nodes). Note that this expansion of the normal distribution function provides the basis for a generalization of the bond orientational order parameter and thus encodes the symmetries expressed by nodal IMDSs.

**Level set Model.** The model of the DD (222) twin boundary was made using a level set representation of the DD[51]. The level set equation used is

F(x,y,z) = sin(2πx)×sin(2πy) sin(2πz) + sin(2πx)×cos(2πy) cos(2πz) + cos(2πx)×sin(2πy) cos(2πz) + cos(2πx)×cos(2πy) sin(2πz) = t

By setting t to ±0.695, The IMDS of DD structure with each IMDS enclosed ~21% volume can be written. The surface file written by the level set equation was introduced into Avizo software and the surface was converted into solid network volume and further sliced into 2D image stack for the construction of twin boundary model.

**Author contributions**

X.F. and E.L.T. conceived the concept and designed the experiments. X.F. performed the experiments and all authors analyzed the data and participated in writing the manuscript.

**Competing interests**

The authors declare no competing interests.

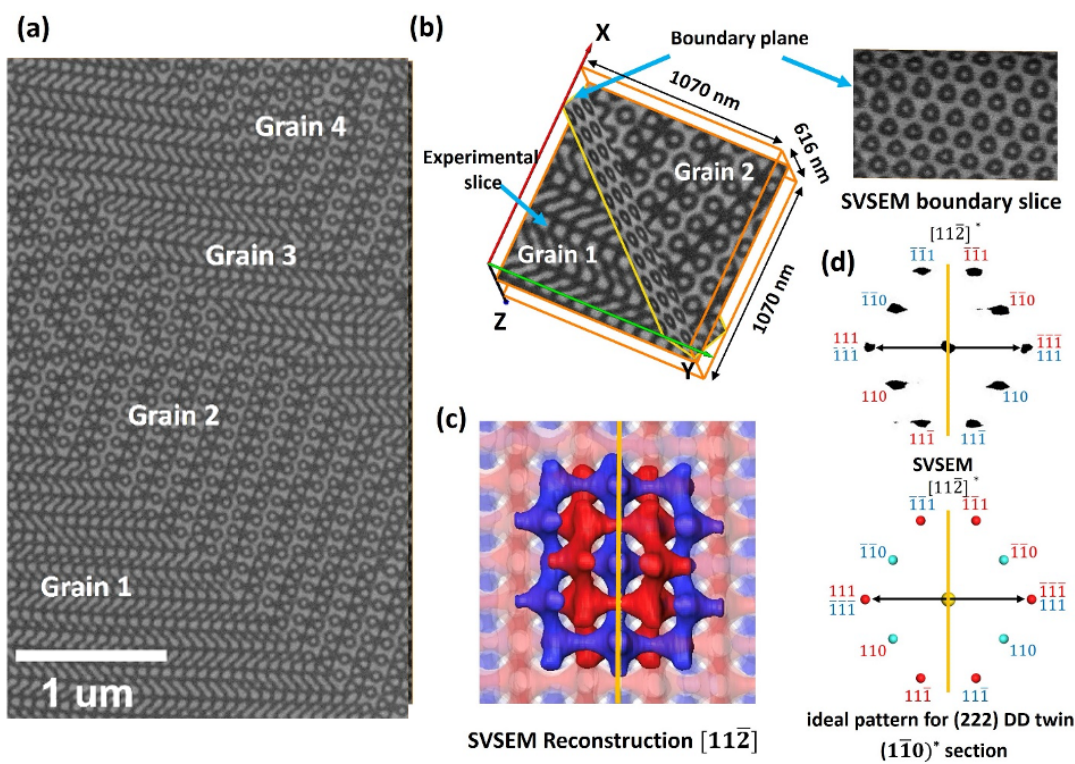

**Figure 1. Identification of TBs in PS-PDMS BCP.** (a) A low magnification SEM image showing four adjacent grains with distinct 2D patterns separated by sharp boundaries. (b) Perspective view of the SVSEM 3D data in the laboratory coordinate frame with the boundary plane viewed at an oblique angle and (right) viewed along the [111] normal. The experimental 2D SEM images are collected parallel to the X-Y plane and the FIB slicing direction is along Z. (c) SVSEM reconstruction of the interpenetrating PDMS networks spanning the TB viewed along the [11$\bar{2}$]. The highlighted partial reconstruction is embedded within surrounding translucent networks with the TB plane indicated by the orange line. (d) (top) (1$\bar{1}$0)$^*$ section of 3D FFT pattern of TB volume SVSEM reconstruction (shown in Figure S3c) compared with (bottom) corresponding pattern for an ideal cubic DD TB. Orange line indicates [11$\bar{2}$]$^*$ reciprocal vector. From the positions of the Bragg peaks, we find that the respective unit cell of each grain (see Table S1 in Supporting Information) is mildly distorted from cubic, likely due to shrinkage stresses from solvent evaporation during solution casting, as often noted in solution-cast BCP materials[17].

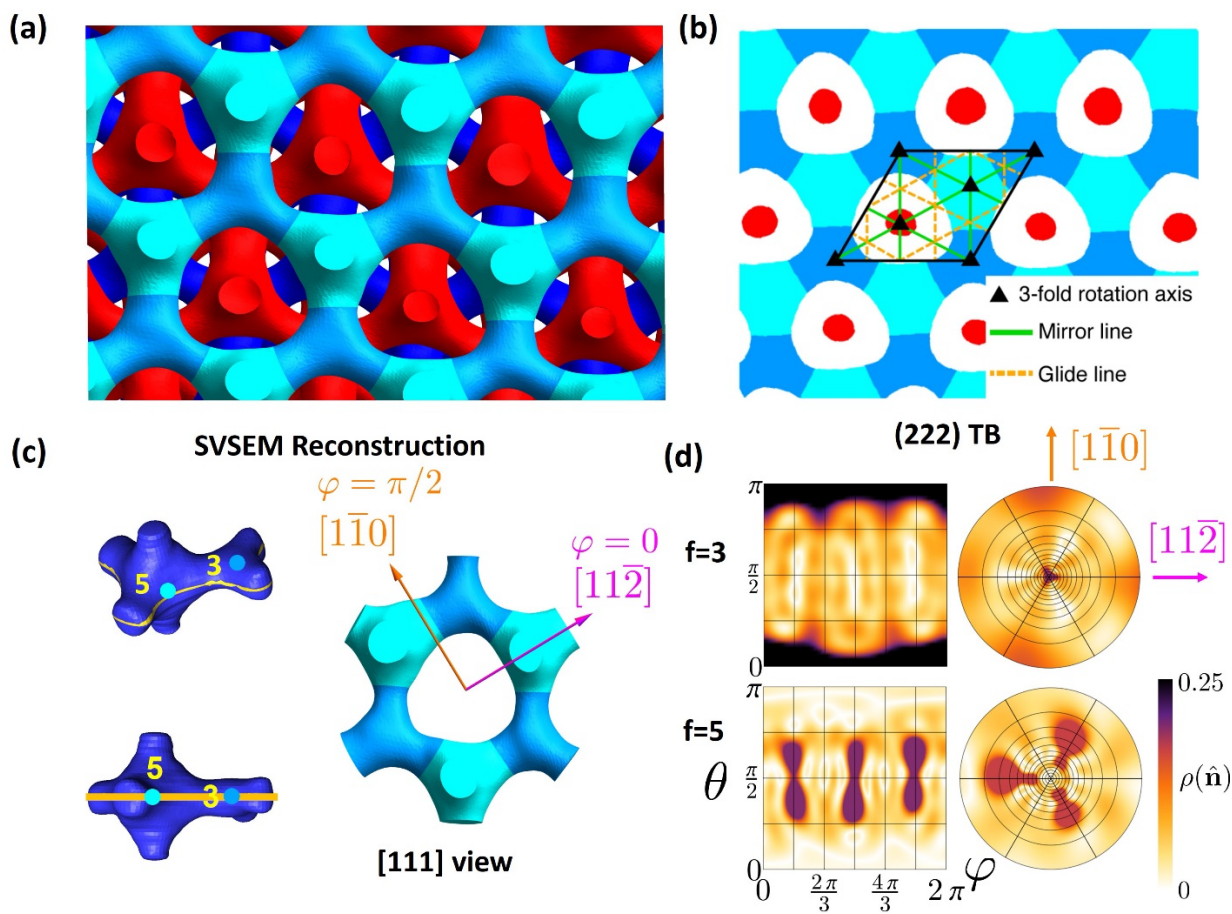

**Figure 2. Visualizing the (222) DD TB boundary nodes and symmetry analysis.** (a) Visualization of reconstructed PDMS networks (blue and red tetrahedral networks) and the frontmost TB plane region with cyan and light blue nodes. The point group symmetry of both the f = 5 (cyan) and f = 3 (light blue) boundary nodes is clearly $\bar{6}m2$. (b) 2D plane symmetry elements associated with the (222) DD TB (2D slice from the SVSEM reconstruction). The (222) boundary plane group symmetry is p3m1 with the symmetry elements labeled for one unit cell. The f = 3 nodes are in Wyckoff site 1a, the f = 5 are in 1b and the red network passes through site 1c. (c) Two perspective views of the blue TB network showing nodes with functionality of 5 and 3. An enlarged view of a 6-(5,3)$^3$ loop in the boundary plane showing the alternating node types. (d) Heat maps of the IMDS normal distribution function averaged over all f=3 (top) and f=5 (bottom) nodes in the TB. These distributions show both symmetry about the equator ($\theta$, $\varphi$ chart on left) as well as 120-degree rotational symmetry about [111] (stereographic projection, right), consistent with $\bar{6}m2$ symmetry.

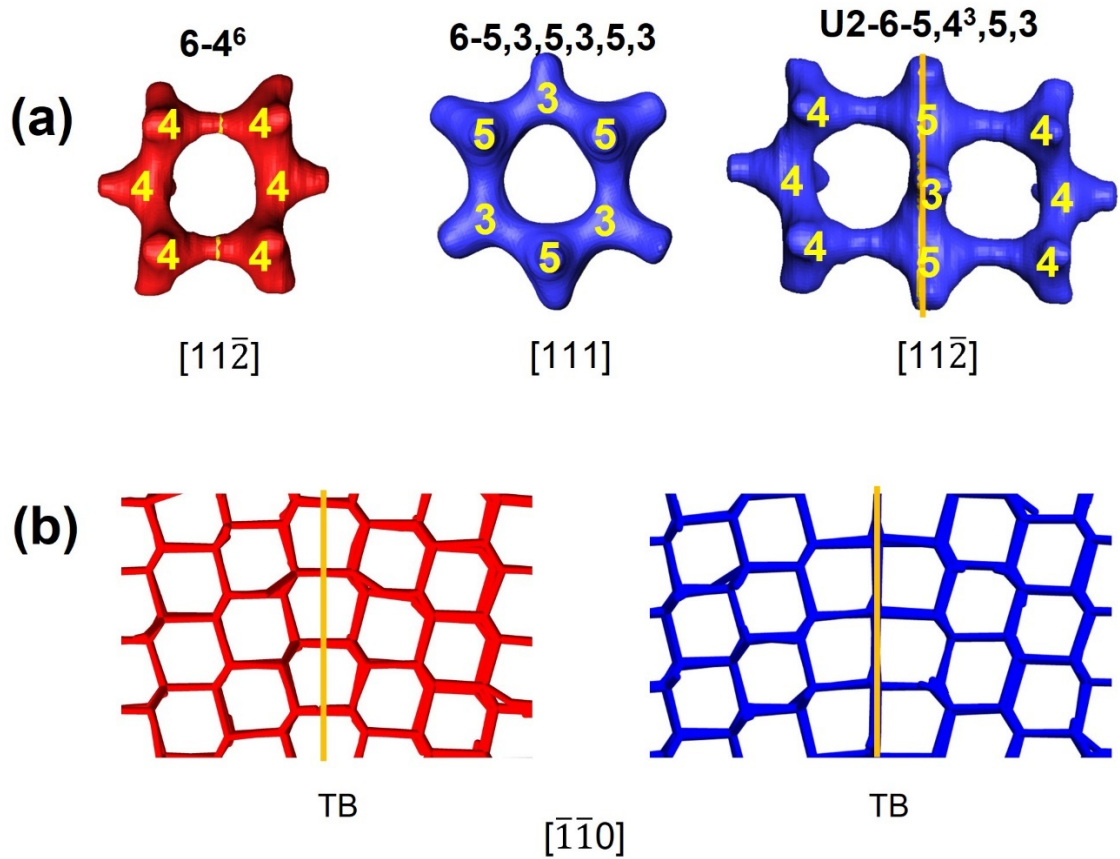

**Figure 3. Geometry and topology of the TB in DD.** (a) SVSEM reconstructions of the three types of boundary loops. In the red network, no boundary nodes are created while two new types of boundary nodes (f = 5 and f = 3) are created in the blue network. (b) Skeletal graph of the red network displaying the set of parallel struts that cross normal to the boundary; and skeletal graph of the blue network indicating that this PDMS network coincides with the plane of the boundary. Consequently, the network reorganizes into a 2D periodic array of 6-(5,3)³ loops. The interpenetrating nature of the blue and red networks is best understood from viewing Figures 1c and 2a.

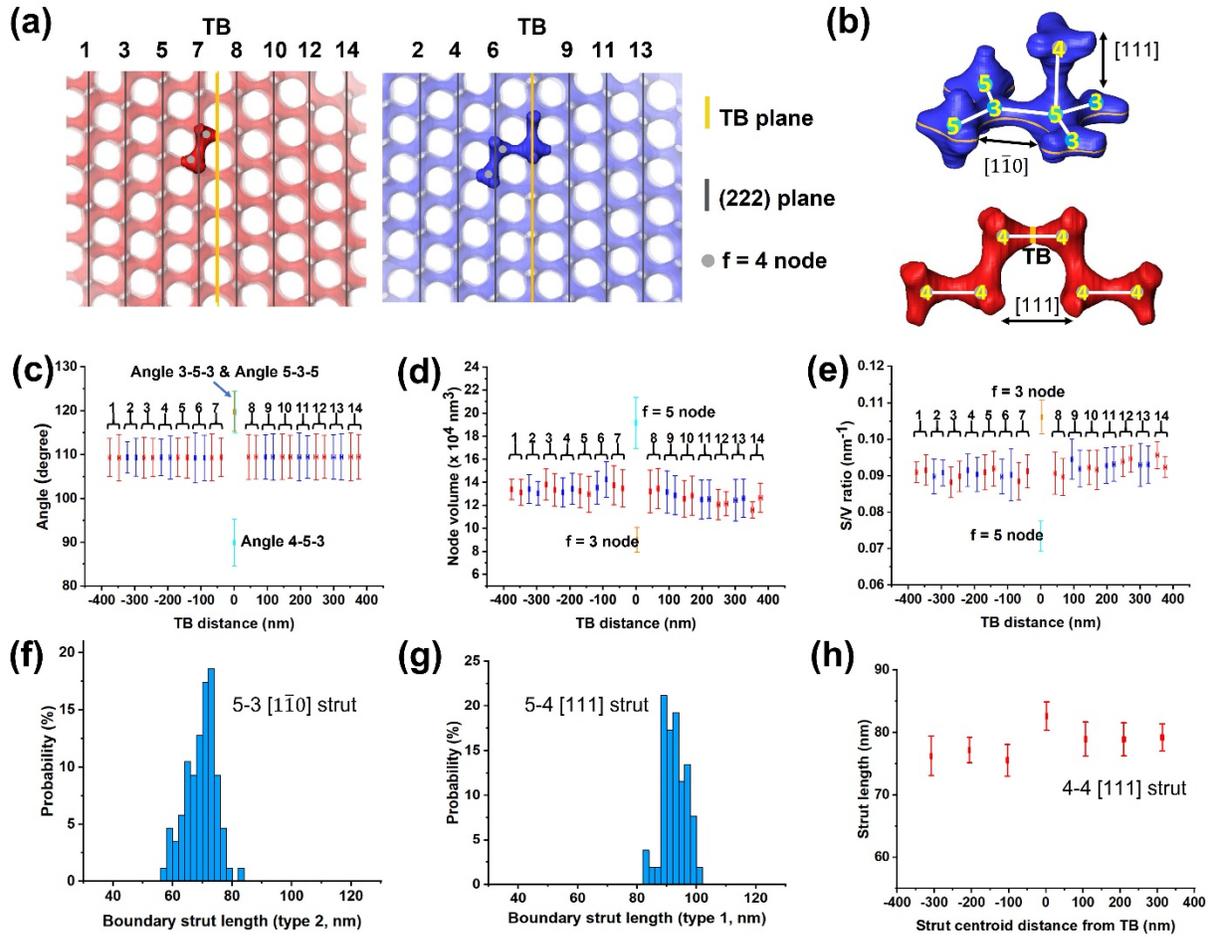

**Figure 4. Malleability for accommodation of the TB.** (a) IMDS structures of experimental SVSEM reconstruction showing positions of various structural features in two layers each of the red and blue networks along the perpendicular from the (222) TB. (b) IMDS based on SVSEM data showing new boundary nodes and boundary struts in the blue and red networks. In the blue network, for the boundary strut length, there are 2 types: 5-3 strut along <110> direction and 5-4 strut along [111] direction; for the boundary internode angle, there are 2 types for $f = 5$ nodes: Angle 4-5-3 = 90°, Angle 3-5-3=120°; and one type for $f = 3$ nodes: Angle 5-3-5=120°. In the red network, there is no change in node connectivity, the boundary strut is perpendicular to the TB oriented along the [111] direction. (c) Internode angle distribution of nodes in different groups versus the distance between nodes and TB plane (for each data point, at least 78 angles are measured). (d) Node volume distribution in different groups versus the distance between nodes and TB plane (for each data point, at least 6 nodes are measured). (e) Node surface to volume ratio distribution in different groups versus the distance between nodes and TB plane (for each data point, at least 6 nodes are measured). Distribution of strut lengths in and normal to the TB: (f) blue network boundary $f = 5$ node to off boundary $f = 4$ node length distribution measured from 52 struts; (g) blue network $f = 5$ node to $f = 3$ node boundary strut length distribution measured from 86 struts; (h) red network strut length distribution of $f = 4$ node to $f = 4$ node struts which are perpendicular to TB plane versus the distance between the strut centroid and the TB plane (for each data point, at least 7 struts are measured). Error bars represent a standard deviation about the averaged value.

# Block Copolymer Double Diamond Twin

Xueyan Feng[1,2,*], Michael S. Dimitriyev[3] and Edwin L. Thomas[1,*]

1. Department of Materials Science and Engineering, Texas A&M University, College Station, TX, USA

2. State Key Laboratory of Molecular Engineering of Polymers, Department of Macromolecular Science, Fudan University, Shanghai, China

3. Department of Polymer Science and Engineering, University of Massachusetts, Amherst, MA, USA

*E-mail: fengxueyan@fudan.edu.cn, elt@tamu.edu

Table S1.   Unit cell parameters of DD unit cells based on 3D FFT.

|         | a     | b     | c     | α   | β   | γ   |
|---------|-------|-------|-------|-----|-----|-----|
| **Grain 1** | 89 nm | 81 nm | 90 nm | 88° | 89° | 86° |
| **Grain 2** | 85 nm | 86 nm | 89 nm | 88° | 91° | 84° |

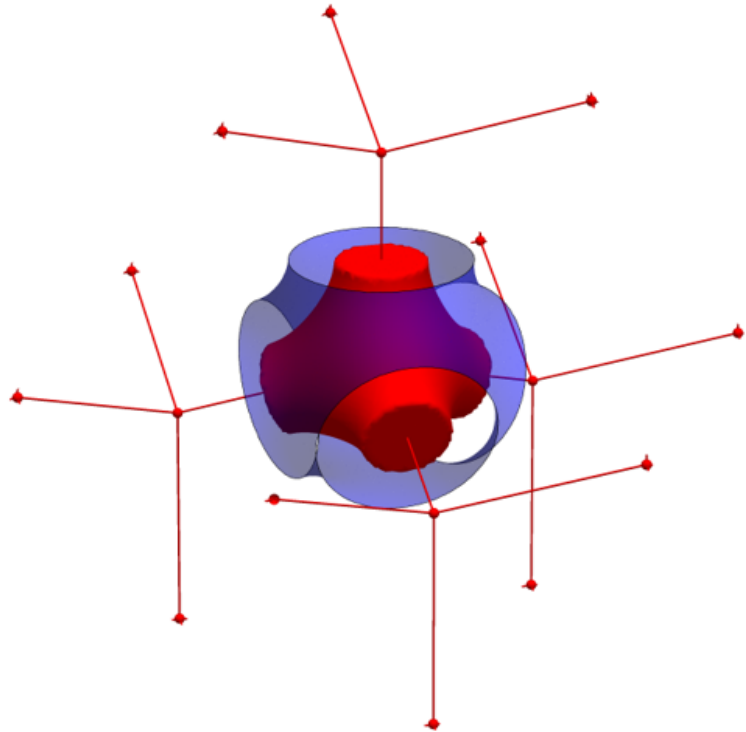

**Figure S1. Model of basic motif (i.e., f = 4 node or mesoatom) of the DD networks.** The intermaterial dividing surface (IMDS) is rendered in red, and the triply periodic minimal surface (TPMS) is rendered in blue. The skeleton of the minority component domain is shown as red sticks. The basic motif occupies Wyckoff site 2a with site symmetry $\bar{4}3m$ in space group $Pn\bar{3}m$.

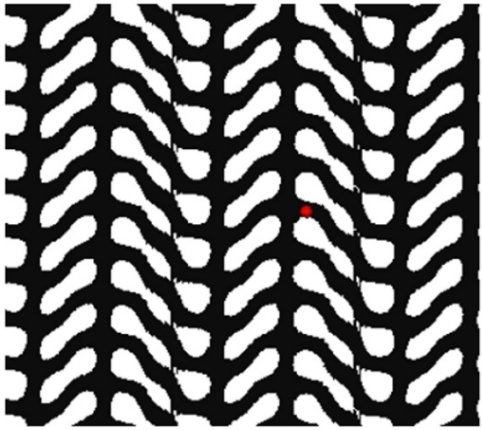 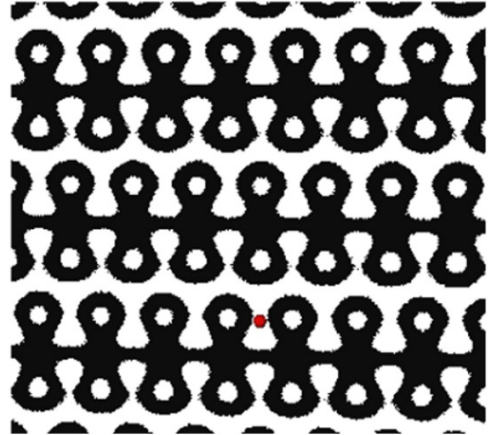

{410} slice         {223} slice

**Figure S2.** 2D slices from level set DD model showing similar patterns as in the experimental 2D SEM image in Figure 1a.

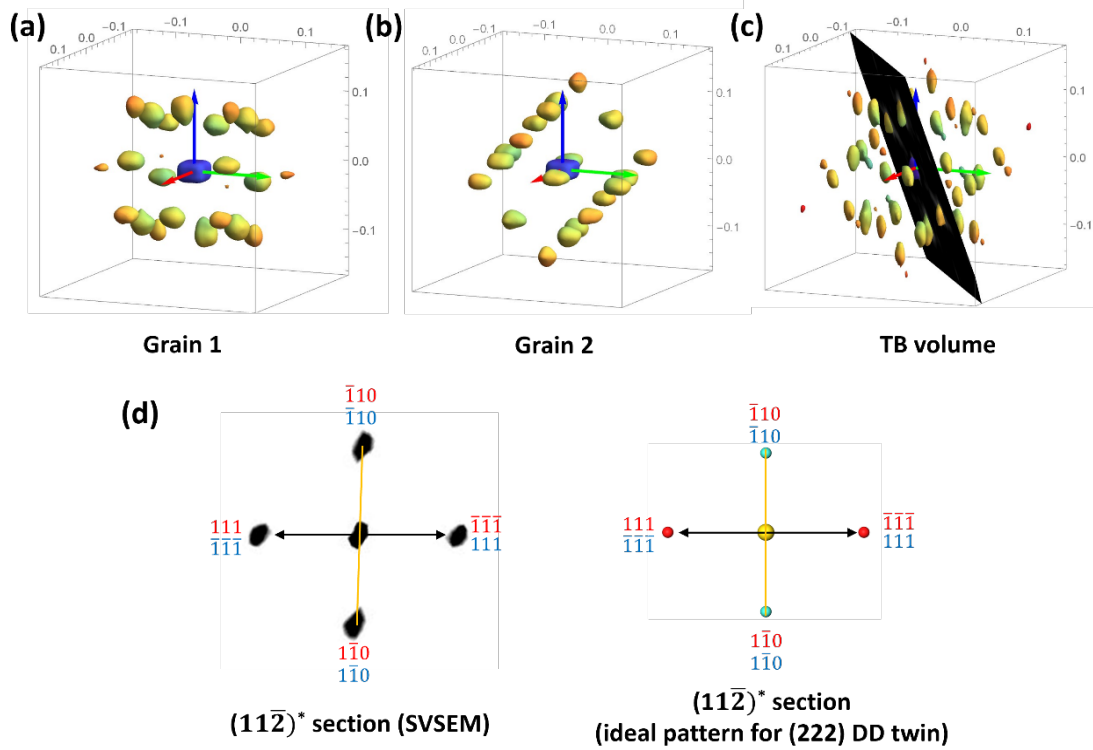

**Figure S3.** 3D FFT patterns of SVSEM reconstructions of grain 1 (a), grain 2 (b), and a larger volume containing portions of grain 1, grain 2 and the TB region (c). The inclined black plane is an example of 2D section of the 3D FFT pattern. (d) $(11\bar{2})^*$ section of 3D FFT pattern of SVSEM reconstructed TB volume compared with corresponding ideal cubic DD TB pattern. Orange line indicates $[1\bar{1}0]^*$ reciprocal vector direction, which lies in the TB plane. 3D FFT patterns in (a) and (b) are based on volume with ~145 unit cells, and 3D FFT pattern in (c) is based on volume with ~1090 unit cells. The FFT spots in (a-c) are highlighted as quasi-spheriodal volumes, colored according to distance from the origin (blue to red).

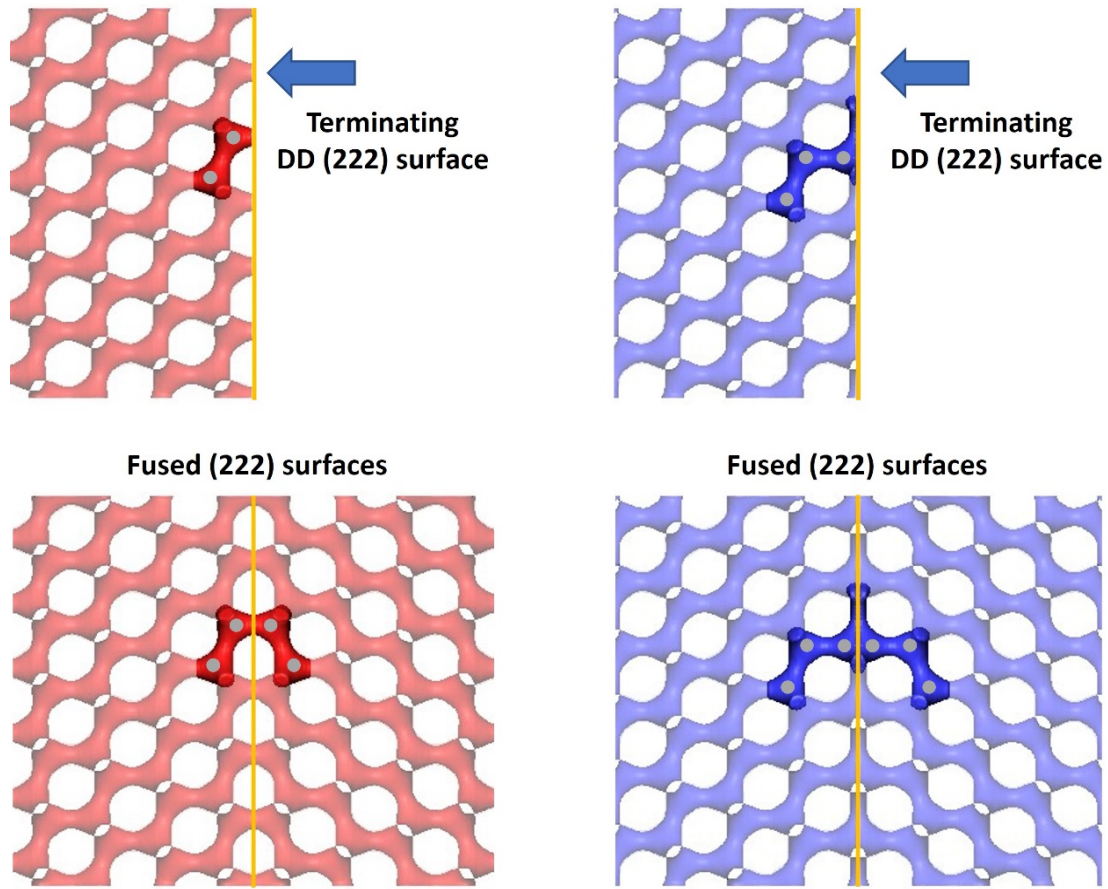

**Figure S4. Construction of the level set model of the (222) TB.** (a) The level set double diamond model for 42 volume % PDMS networks is cut to provide a (222) terminating surface for the red and blue networks (nodes are shown as grey dots). The terminating plane passes through the center of struts in the red network but is offset from the tetrahedral nodes in the blue network. To form the TB level set model, the terminated 3D model is mirrored and the two portions are joined side by side at the (222) surface without further adjustment.

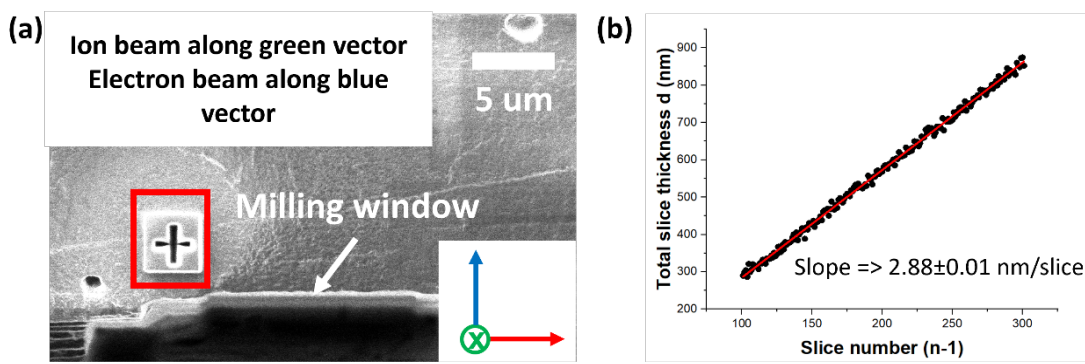

**Figure S5. Monitoring of experimental slice thickness**. (a) Fiducial (within red square) in ion beam view for the registration of FIB slicing and slice thickness monitoring during SVSEM experiments. (b) Slice thickness monitoring: the distance between the milling surface of the n$^{th}$ slice and the milling surface of the first slice (total slice thickness d, slices are registered with the fiducial) was measured based on FIB images, and then d vs slice number (n-1) was plotted and shows a linear relationship with a slope of 2.88±0.01, which is the averaged slice thickness in nm. The resolution of the 3D reconstruction is ~(3nm)$^3$, allowing good distinction of features >12-15 nm, more than adequate for the length scale of the tubular networks examined.

**Prior Imaging of Diamond Twins in Soft Matter**

Recent studies using TEM and electron diffraction have identified the presence of twins in cubosome particles and characterized the structure of the TB. Han et al.[1,2] investigated porous silica cubosomes derived by calcination of a 5 component – water-based precursor. The porous silica-air structure takes on a 3D shape that can be approximated by a pair of parallel surfaces extending from Schwarz's D triply periodic minimal surface (TPMS) [3,4]. A 2D TEM image containing a sharp linear boundary feature was compared to simulated images based on the TPMS D surface incorporating a {111} type twin. The best agreement was for a laterally-shifted TB plane located at (111)+0.5 (equivalent to a (222) plane in the $Pn\bar{3}m$ structure), with D surface grains joined by a layer resembling Schwarz's H surface. Based on modeling, a skeletal graph was constructed for each air labyrinth with one air channel network exhibiting new boundary nodes (i.e. f = 5 nodes). Twins[5] are also found in amphiphilic cubosome nanoparticles made by self-assembly of polyoxometate units linked to polyhedral oligomeric silsesquioxane units are also thought to be (111) twins. The specimen was imaged using high angle annular dark field scanning TEM and the 2D projections of the twin fit with a level-set model approximating the (111) twinned TPMS D surface. There was special interest in the continuity of air channels across the TB since this is key for catalytic applications. The twin boundary altered the tetrahedral point group symmetries to $D_{3h}$ ($\bar{6}m2$) and provided channel connectivity across the TB. However, as in the previous studies, TEM projections for specimens having unit cells with lattice parameters of ~ 15nm across samples of 600 nm thickness cannot provide direct visualization of the detailed 3D structure of the TB. SEM images of polystyrene-*b*-poly(ethylene oxide) assembled DD polyhedral cubosomes also suggest that the particles might be multiply twinned.[6]